\documentclass[prl, twocolumn,amsmath,amssymb, superscriptaddress, nofootinbib]{revtex4-1}
\usepackage{graphicx}% Include figure files
\usepackage{natbib,hyperref}
\usepackage{amsmath,epsfig,color,amssymb}
\usepackage{bibunits}

\defaultbibliography{bibfqhe,graphene_hafezi}
\defaultbibliographystyle{apsrev4-1}

 \newcommand{\ket}[1]{\left|#1\right\rangle} %|"cosa">
 \newcommand{\bra}[1]{\left\langle#1\right|} %<"cosa"|
 \newcommand{\braket}[2]{\left< #1 \vphantom{#2} \right|
  \left. #2 \vphantom{#1} \right>} % for Dirac brackets
  
  \definecolor{new}{rgb}{.08,.05,.8}

\begin{document}
\begin{bibunit}

\author{Areg Ghazaryan}
\affiliation{Department of Physics, City College, City University of New York, New York, NY 10031, USA}
\author{Tobias Gra\ss}
\affiliation{Joint Quantum Institute, NIST and University of Maryland, College Park, MD 20742, USA}
\affiliation{Department of Physics, College Park, MD 20742, USA}
\author{Michael J. Gullans}
\affiliation{Joint Quantum Institute, NIST and University of Maryland, College Park, MD 20742, USA}
\affiliation{Joint Center for Quantum Information and Computer Science, University of Maryland, College Park, Maryland 20742, USA}
\author{Pouyan Ghaemi}
\affiliation{Department of Physics, City College, City University of New York, New York, NY 10031, USA}
\affiliation{Department of Physics, Grad. Center, City University of New York, New York, NY 10016, USA}
\author{Mohammad Hafezi}
\affiliation{Joint Quantum Institute, NIST and University of Maryland, College Park, MD 20742, USA}
\affiliation{Department of Electrical Engineering and IREAP, University of Maryland, College Park, MD 20742, USA}
\affiliation{Department of Physics, College Park, MD 20742, USA}
\title{Light-induced fractional quantum Hall phases in graphene}

\begin{abstract}
We show how to realize two-component fractional quantum Hall phases in monolayer graphene by optically driving the system. A laser is tuned into resonance between two Landau levels, giving rise to an effective tunneling between these two synthetic layers. Remarkably, because of this coupling, the interlayer interaction at non-zero relative angular momentum can become dominant, resembling a hollow-core pseudo-potential.  In the weak tunneling regime, this interaction favors the formation of singlet states, as we explicitly show by numerical diagonalization, at fillings $\nu=1/2$ and $\nu=2/3$. We discuss possible candidate phases, including the Haldane-Rezayi phase, the interlayer Pfaffian phase, and a Fibonacci phase. This demonstrates that our method may pave the way towards the realization of non-Abelian phases, as well as the control of topological phase transitions, in graphene quantum Hall systems using optical fields and integrated photonic structures.
\end{abstract}

\maketitle

\textit{Introduction.} The fractional quantum Hall (FQH) effect is a fascinating phenomenon, where electron-electron interactions and a magnetic field lead to strong correlations \cite{ChakrabortyBook,PrangeBook,JainBook}. Soon it was realized \cite{Halperin83,Chakraborty84,Zhang84,Xie89} and experimentally confirmed \cite{Eisenstein89,Eisenstein90} that the electron spin plays an important role at several fillings.  More generally, multicomponent FQH phases \cite{Girvin97} occur in systems with subbands, as wide quantum wells \cite{Suen94,Lay97,Shabani13,Liu15}, with layers, as double wells \cite{Eisenstein92,Suen92}, or with degenerate valleys, as AlAs quantum well \cite{Bishop07} or graphene \cite{Dean11,Feldman12,Feldman13,zibrov17}. Much effort has been made towards engineering system parameters like tunneling, in order to realize different phases. Here we propose a new method based on light-matter interactions which enables flexible control in a synthetic FQH bilayer.
%However, these approaches can add unwanted side effects, and therefore, it is desirable to investigate other control methods.

Interactions between light and graphene quantum Hall samples have been subject of many theoretical \cite{Furchi12,Hagenmuller12,Chirolli12,Pellegrino14} and experimental \cite{Byszewski:2006jsa, Jiang:2007hm, Orlita:2008bm, Orlita:2011ca} studies. FQH phases in integrated GaAs quantum well-cavity structures have also been explored experimentally \cite{Smolka14}. A distinctive feature of graphene is the linear dispersion, resulting in non-equidistant Landau levels (LLs) \cite{Goerbig11} which can selectively be coupled with resonant light.  

\begin{figure}
\includegraphics[width=8.5cm]{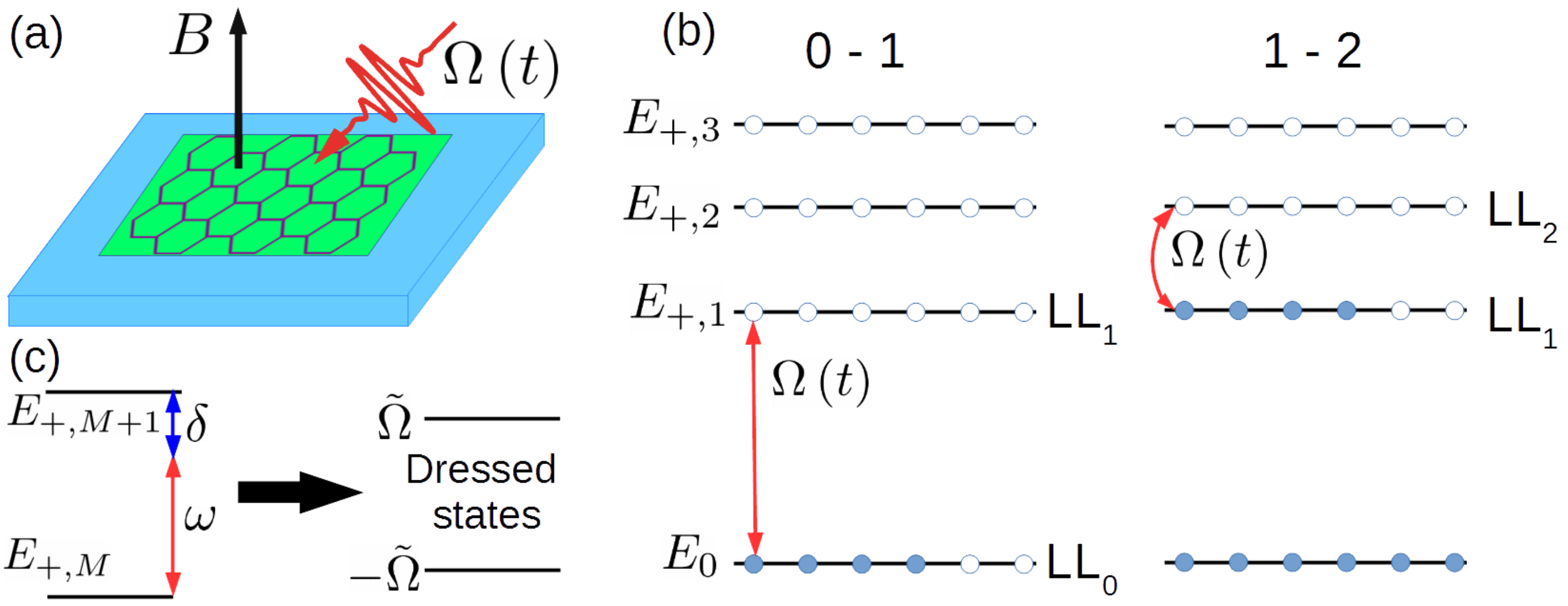}
\caption{\label{fig:schematic} (a) A single graphene layer driven by light at Rabi frequency $\Omega$. (b) LL structure with partial filling and optical transitions  ${\rm LL}_{0-1}$ and ${\rm LL}_{1-2}$. (c) Formation of dressed states due to coupling between two LLs. }
\end{figure}

The present letter explores this possibility. While in the absence of light a large gap freezes out all but one LL, resonant light coupling to an empty level provides an effective tunneling to this new degree of freedom. The coupled LLs can then be viewed as two layers of a physical bilayer. Depending on the tunneling rate, which is tunable via the laser intensity, the system either polarizes in the lower dressed LL, or it realizes a singlet phase. Analysis of the Coulomb interaction between different LLs shows that the repulsion between singlet pairs becomes particularly small when first and second LLs are coupled, resembling a hollow-core Haldane pseudo-potential  \cite{Haldane1990,PrangeBook}. Such interaction favors the formation of a many-body singlet phase, which we confirm explicitly by exact diagonalization (ED), at filling $\nu=1/2$ and $\nu=2/3$. We identify the polarized phases as a composite Fermi sea ($\nu=1/2$) \cite{HLR}, and a quasi-hole conjugate 1/3 Laughlin state ($\nu=2/3$) \cite{laughlin83}. The singlet phase at $\nu=1/2$ has good overlap with the Haldane-Rezayi phase  \cite{HR}, an intriguing gapless quantum Hall phase \cite{green2001,RR96,milovanovic}. Some evidence of non-Abelian quantum Hall singlets are found at $\nu=2/3$, including the Fibonacci phase \cite{vaezi14} and the interlayer Pfaffian phase \cite{ardonne02,maissam10}, which are interesting candidates for topological quantum computing \cite{nayak08}.

\textit{System.}
We consider a monolayer of graphene under a perpendicular magnetic field, in the quantum Hall regime \cite{Goerbig11}.  We restrict ourselves to a single valley, and assume that the electron spin is fully polarized. The single-particle states are given by spinors of the form $\Psi_{\gamma,n,j}(z)=\left(-\gamma C^{-}_{n}\phi_{n-1,j}(z), C^{+}_{n}\phi_{n,j}(z) \right)^T$, where $C^{\pm}_{n}=\sqrt{\left(1\pm\delta^{}_{n,0}\right)/2}$ are coefficients, $z=x-iy$ are spatial coordinates, and $\phi_{n,j}(z)$ are the (gauge-dependent) non-relativistic Landau level (LL) wave functions, characterized by the LL index $n\geq0$, and a second quantum number $j\geq0$ \cite{Goerbig11}.  In symmetric gauge, $j$ specifies the $z$-component of angular momentum, while in Landau gauge, it defines momentum along one direction in the plane. In graphene, a third quantum number $\gamma=\pm$, distinguishes between states at positive and negative energy, $E_{\gamma,n}=\gamma\omega^{}_\mathrm{c}\sqrt{n}$, where $\omega^{}_\mathrm{c}=\sqrt{2}v_\mathrm{F}/l^{}_{B}$ and $l^{}_{B}=\sqrt{ c/eB}$ is the magnetic length. The magnetic field strength is $B$ and the Fermi velocity is $v_\mathrm{F}$. In the following, we drop the index $\gamma$, and assume $\gamma=+1$, without loss of generality.

As illustrated in Fig.~\ref{fig:schematic}(a,b), we consider a coupling between the partially filled $n=M$ level at the Fermi surface to the empty LL $n=M+1$, described by ($\hbar=1$):
\begin{equation}
{\cal H}_\mathrm{coup}=\sum_{j,j^\prime}\Omega_{j,j^\prime}(t)c^\dagger_{M+1,j}c_{M,j^\prime}+\mathrm{H.c.}.
\end{equation}
Here, $c^\dagger_{M,j}$ and $c_{M,j}$ are the creation and annihilation operators in $\mathrm{LL}_M$ with the (angular) momentum quantum number $j$. For simplicity, we assume a plane wave drive, which acts uniformly on all orbitals: $\Omega_{j,j^\prime}(t)=2\Omega\delta_{j,j^\prime}\cos (\omega t)$, with $\omega$ the drive frequency, and the Rabi frequency $\Omega$. Within the rotating frame, transformed to by $U=\exp\left[-\frac{i}{2}\omega t\sum_j\left(c^\dagger_{M,j}c^{}_{M,j}-c^\dagger_{M+1,j}c^{}_{M+1,j}\right)\right]$, a rotating-wave approximation (RWA) removes the time-dependence from the coupling. The effective single-particle Hamiltonian then reads
\begin{equation}
{\cal H}_{\rm sp}=\sum_j -\frac{\delta}{2}\tau_z^{(j)}+\Omega\tau_x^{(j)},
\label{eq:dress}
\end{equation}  
with $\delta$ the detuning of the light from the LL resonance, i.e.~$\delta=E^{}_{M+1}-E^{}_{M}-\omega$.
The notation of Eq.~(\ref{eq:dress}), using Pauli operators $\tau_z^{(j)} \equiv \ket{M,j}\bra{M,j} - \ket{M+1,j}\bra{M+1,j}$, and $\tau_x^{(j)} \equiv \ket{M,j}\bra{M+1,j} + \ket{M+1,j}\bra{M,j}$, captures the analogy to a spin-1/2 system, if the $n$ quantum number is interpreted as the $z$-component of spin, or to a bilayer system if $n$ is associated with a layer index. The first term in Eq.~(\ref{eq:dress}) corresponds to a Zeeman term (in the spin picture), while the second term mimics interlayer tunneling (in the bilayer picture). Both terms are independently tunable. The single-particle eigenstates are dressed LLs at energies $\pm \tilde \Omega = \pm\sqrt{\frac{\delta^2}{4}+\Omega^2}$, see Fig.~\ref{fig:schematic}(c). 
While strong coupling and/or far detuning lead to polarization in the lower dressed level, both manifolds are occupied if the gap between dressed states becomes small compared to the interaction strength, $e^2/\epsilon l_B$, i.e. if $\Omega$ and $\delta$ are sufficiently small. The transition occurs near $\Omega \sim 10^{-2}$ (in units of $e^2/\hbar \epsilon l_B$), above the threshold required for thermalization in the rotating frame Hamiltonian, $\Omega >10^{-4} $, as estimated below.

Applying RWA  to the interactions, the many-body Hamiltonian reads ${\cal H}={\cal H}_{\rm sp} + {\cal H}_{\rm int}$, where
%\begin{align}
%{\cal H}_{\rm int} =& \sum_{\substack{n_1,n_2\\n_3,n_4}}\sum_{\substack{j_1,j_2\\j_3,j_4}}A_{\substack{n_1,j_1,n_2,j_2\\n_3,j_3,n_4,j_4}}\delta_{n_1+n_2,n_3+n_4}
%\times \nonumber \\ &
%c^{\dagger}_{n_1,j_1}c^{\dagger}_{n_2,j_2} c_{n_3,j_3}c_{n_4,j_4},
%\label{fullHam}
%\end{align}
\begin{align}
{\cal H}_{\rm int} =& \sum_{\{n,j\}}A^{n_1,j_1,n_2,j_2}_{n_3,j_3,n_4,j_4}\delta_{n_1+n_2,n_3+n_4}
c^{\dagger}_{n_1,j_1}c^{\dagger}_{n_2,j_2} c_{n_3,j_3}c_{n_4,j_4}.
\label{fullHam}
\end{align}
The interaction matrix elements $A^{n_1,j_1,n_2,j_2}_{n_3,j_3,n_4,j_4}$ are the same as without light, but the RWA enforces conservation of single-particle energy, i.e. $\delta_{n_1+n_2,n_3+n_4}$. 

\begin{figure}
\includegraphics[width=8.5cm]{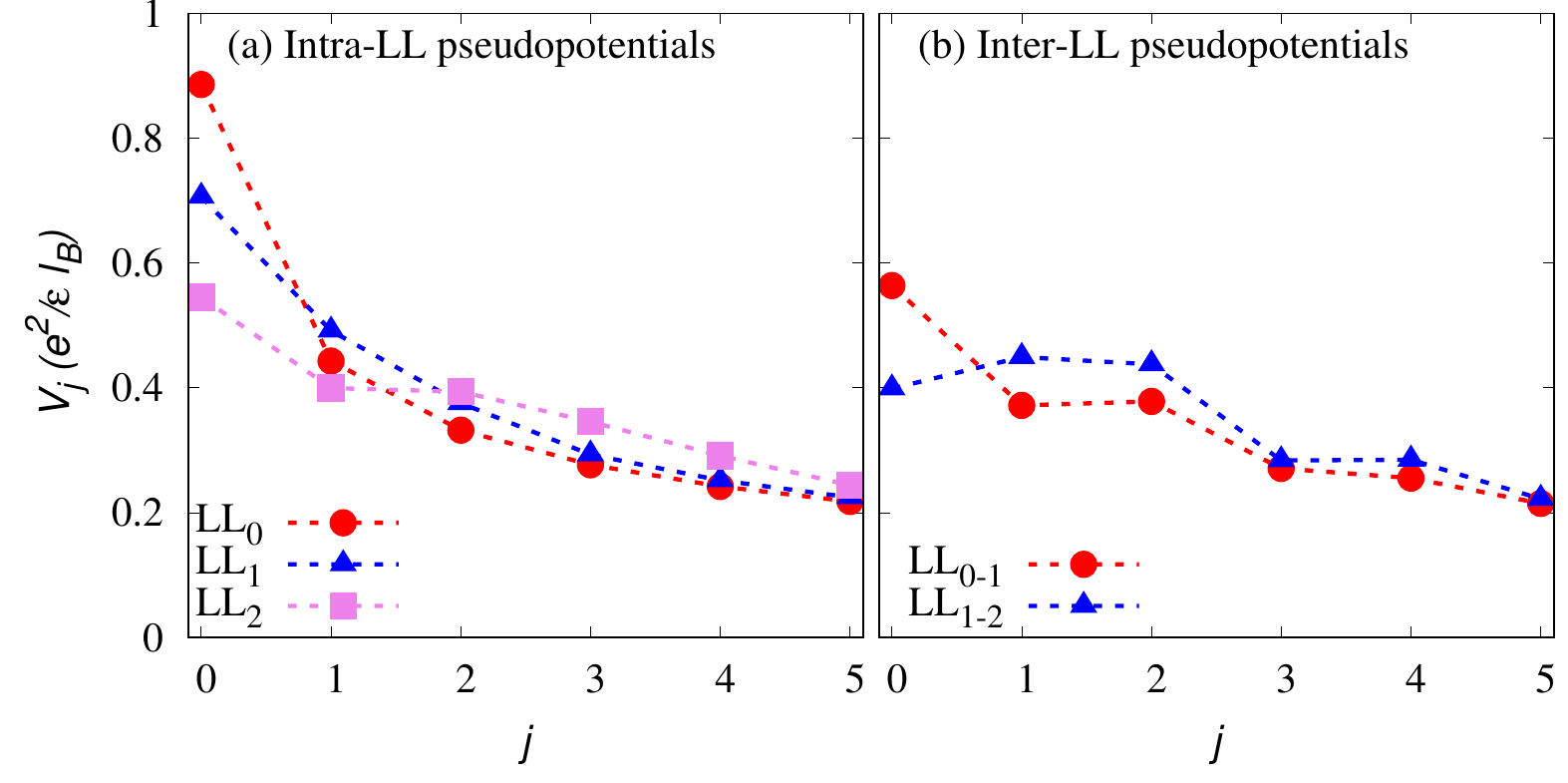}
\caption{\label{fig:ppots} (a) Pseudopotentials for scattering of two particles in the same graphene LL, $n=0, \ n=1,$ and $n=2$. (b) Pseudopotentials for scattering in different LLs, as defined in Eq. (\ref{vinter}). If $n=1$ is coupled to $n=2$, $V^{\rm inter}_j$ is dominated by the contribution $j=1$. 
}
\end{figure}

\textit{Results.}
Before numerically solving ${\cal H}$ for small systems, we gain some intuition by decomposing the interactions into Haldane pseudopotentials \cite{Haldane1990}. These pseudopotentials describe the interaction strength $V_j$ of two particles at fixed relative angular momentum $j$. In our case, we distinguish between intra-layer processes $V_j^{(n)}$ within ${\rm LL}_n$, and inter-layer processes, $V_j^{\uparrow \downarrow,\downarrow \uparrow}$ and $V_j^{\uparrow \downarrow,\uparrow \downarrow}$, where the index $\uparrow (\downarrow)$ shall denote the ${\rm LL}_{M+1}$ (${\rm LL}_M$). 
Clearly, the difference between $V_j^{(M+1)}$ and $V_j^{(M)}$ breaks the $\mathbb{Z}_2$ symmetry usually present in a system of two equivalent layers. However, as seen from Fig.~\ref{fig:ppots}(a), this breaking is weak, since only potentials at odd $j$ contribute to the intra-LL scattering of fermions, whereas the strongest $n$-dependence occurs for $V_0^{(n)}$. A more important difference to standard bilayer systems stems from the interaction $V_j^{\uparrow \downarrow,\uparrow \downarrow}$ where scattering particles exchange their LL index, while in standard bilayers only density-density-type interactions $V_j^{\uparrow \downarrow,\downarrow \uparrow}$ occur between two layers. Both types of inter-LL processes can conviently be accounted for by a single pseudopotential $V^{\rm inter}_j$. Therefore, we switch to a singlet/triplet basis, $\ket{\pm} \sim \ket{\uparrow\downarrow} \pm \ket{\downarrow\uparrow}$, where the corresponding pseudopotentials are $V^\pm_j = (V_j^{\uparrow \downarrow,\downarrow \uparrow} \pm V_j^{\uparrow \downarrow,\uparrow \downarrow})/2$. Since $\ket{+}$ ($\ket{-}$) is even (odd) under particle exchange, it requires odd (even) $j$, and it is sufficient to consider
\begin{align}
 V^{\rm inter}_j = \left[V_j^{\uparrow \downarrow,\downarrow \uparrow} + (-1)^j V_j^{\uparrow \downarrow,\uparrow \downarrow}\right]/2.
 \label{vinter}
\end{align}
As seen from Fig.~\ref{fig:ppots}(b), these inter-LL pseudopotentials $V^{\rm inter}_j$ are dominated by $j=0$ for a coupling between ${\rm LL}_0$ and ${\rm LL}_1$ (denoted ${\rm LL}_{0-1}$). In contrast, the repulsion between singlets at $j=0$ is suppressed for a coupling between ${\rm LL}_1$ and ${\rm LL}_2$ (denoted ${\rm LL}_{1-2}$), and $V_1^{\rm inter}$ becomes the dominant contribution. This behavior leads to the general expectation that coupling ${\rm LL}_{1-2}$ favors singlet phases and could give rise to bilayer quantum Hall phases which are derived from a hollow-core Hamiltonian. In the following, we will test this expectation at filling factors $\nu=1/2$ and $\nu=2/3$ using ED on torus \cite{ChakrabortyBook,Haldane85}, sphere, and disk. 
%In the absence of light, the FQH phases in graphene are well studied \cite{ChakrabortyChapter13,Chakraborty13,Goerbig07,Papic09,Toke07} both in single-component and multicomponent regimes. In the calculations, we take the unit of the energy to be $e^2/\epsilon l^{}_B$. All calculations are performed for non-zero detuning ($\delta=0.02$) to rule out singular behavior associated with driving exactly on resonance. Due to the computational complexity, we disregard LL mixing to other levels not involved in the optical transition, which is a good approximation for graphene placed on a substrate \cite{Peterson13,Peterson14}.
%Note that in conventional bilayer systems the layer index acts as a pseudo-spin and only direct type interactions between the layers are present. Due to the fact that in conventional bilayer systems the exchange type interaction is only present for intra-layer interactions, it was initially supposed that FQHE phases should be pseudo-spin polarized, which was proven to be wrong by exact-diagonalization \cite{Chakraborty84,Zhang84,Xie89} and experimental studies \cite{Eisenstein89,Eisenstein90}. The presence of inter-layer cross interaction terms in our case nominally makes the competition between pseudo-spin polarized and unpolarized phases even more important, but it competes with the SU(2) symmetry breaking. The competition between the SU(2) symmetry breaking and inter LL cross interaction terms will be analyzed below.

\paragraph{$\nu=1/2$.} Since the discovery of FQHE understanding the physics of a half filled LL has been a challenge. Early generalizations of the Laughlin wave functions to systems with spin provide an Abelian spin singlet state at $\nu=1/2$, known as the $(331)$-Halperin state \cite{Halperin83}. However, in most systems, no quantum Hall plateaux are observed at $\nu=1/2$. This fact has been explained by Halperin, Lee, and Read through a theory which attaches all magnetic fluxes to composite fermions \cite{HLR}. As a consequence, these fermions do not feel a magnetic field, and may form a compressible Fermi liquid. In an alternative scenario, the composite fermions undergo BCS pairing which, due to the Meissner effect, leads to incompressibility \cite{moore_read,green2001}. The most prominent paired state is the Moore-Read Pfaffian state. It involves $p$-wave pairing, and is spin polarized. In contrast, a spin singlet state can be obtained via $d$-wave pairing, and is known as the Haldane-Rezayi (HR) state \cite{HR}. Evidences of non-Abelian excitations have been discussed for both states \cite{RR96}. The HR phase has been identified as a critical phase between strong and weak pairing \cite{green2001}, providing an example for a gapless FQH system. Hollow-core two-body interactions, i.e. pseudopotentials given by $V^{\rm intra}_j \sim \delta_{j,1}$ and $V^{\rm inter}_j \sim \delta_{j,1}$, yield a parent Hamiltonian for the HR state.

Accordingly, given the pseudopotential structure of coupled LLs discussed above, the HR phase becomes a likely candidate for coupling  ${\rm LL}_{1-2}$. Indeed, for sufficiently weak Rabi frequencies, numerical results support this expectation: In all three geometries, the ground state is a singlet, having large overlaps with the HR state (see Table \ref{table}).  We have also evaluated the overlap with the Jain singlet, which is known to have a large overlap with the ground state of pseudopotential  $V_0 \simeq V_1$ \cite{Moran12}. However, since this overlap decreases rapidly with the system size,  we excluded the Jain singlet as a possible candidate (see Supplementary Material). For the observed singlet phase, the topological degeneracy on the torus is $4q$-fold with ground states at high-symmetry points ${\bf K}=(0,0)$, ${\bf K}=(0,N/2)$, ${\bf K}=(N/2,0)$, and ${\bf K}=(N/2,N/2)$. While this is compatible with a $(331)$-phase, no sizable overlap with this phase are found in any geometry. The HR phase, as obtained from the hollow-core model, exhibits ground states at the same high-symmetry $\bf K$-points, but has two linearly independent ground states ${\bf K}=(0,0)$. This $5q$-fold degeneracy of the HR phase has been discussed as a consequence of its criticality \cite{RR96,green2001}, leading to a zero mode which can be either occupied or empty. 
However, the torus degeneracy of the HR state in the hollow-core model differs from the number of sectors in the underlying conformal field theory which is $4q$  \cite{milovanovic}, suggesting that the fifth ground state is not crucial for realizing the HR phase.  In light of this point and based on the strong numerical evidence, the HR phase appears as the likely description of the observed singlet phase.

%, despite the missing fifth ground state, the singlet is described by the HR phase.

Upon increasing the Rabi frequency, a crossing of energy levels indicates a second-order phase transition (at $\Omega\approx 0.025$ and $\delta=0.02$ in units $e^2/\epsilon l_B$, for $N=8$ electrons on the torus).   The ground state on the strong-coupling side is fully polarized in one LL, and the system exhibits Fermi sea behavior, indicated by ground states at finite angular momentum on the sphere, and at non-zero pseudomomenta on the torus. A Fermi liquid phase is also found for coupling ${\rm LL}_{0-1}$ where this behavior extends to $\Omega \rightarrow 0$.  For ${\rm LL}_{0-1}$, increasing $\Omega$ only rotates the LL polarization from $\langle \sum_j \tau_z^{(j)} \rangle = N$ and $\langle \sum_j \tau_x^{(j)} \rangle = 0$ for  $\Omega \rightarrow 0$, to $\langle \sum_j \tau_z^{(j)} \rangle = 0$ and $\langle \sum_j \tau_x^{(j)} \rangle = N$ for  $\Omega \rightarrow \infty$.
This pseudospin rotation is understood on the single-particle level by assuming that the ground state always remains polarized in the lower dressed LL.

\begin{table}[]
\centering

\begin{tabular}{c||c|c|c}
 &  Sphere  & Disk  & Torus  \\
\hline
$\nu=1/2$ & 0.85 ($N=6$)  &  0.97  & 0.83 $({\bf K}={\bf 0})$ \\
(HR) & 0.75 ($N=8$) &   ($N=6,L=24$)  & 0.72 $({\bf K}\neq{\bf 0})$  \\
& 0.72 ($N=10$) & & ($N=8$) \\
\hline
$\nu=2/3$   & 0.99 ($N=4$) & 0.81 ($N=6,L=18$)& \\
(IP) &  0.55 ($N=8$) &   0.63 ($N=8,L=36$)& \\
 &  0.39 ($N=12$) & & \\
\end{tabular}
\caption{Overlaps of ground states in different geometries, for weak ${\rm LL}_{1-2}$ coupling ($\Omega = 10^{-3}$ and $\delta=0.02$), with HR state ($\nu=1/2$), and with interlayer Pfaffian (IP) state ($\nu=2/3$).  At $\nu=2/3$, fast decay of the overlap with $N$ suggests a different phase, possibly a Fibonacci phase (see discussion), however we are not aware of unique trial wave functions to test the overlaps with this phase.}
\label{table}
\end{table}

\paragraph{$\nu=2/3$.} At filling fractions $1/q$ with $q$ odd, electrons can anti-correlate by forming a Laughlin state \cite{laughlin83}. Similarly, a Laughlin state of holes provides a good trial wave function at $\nu=1-1/q$, including $\nu=2/3$. In a bilayer at $\nu=2/3$, various singlet phases compete with the polarized Laughlin state.
Similar to the $\nu=1/2$ case, Halperin $(mmn)$-states \cite{Halperin83} are possible, including the $(221)$-state and the $(330)$-state, the latter being two uncorrelated copies of the 1/3 Laughlin states. Apart from these Abelian phases, there are also different non-Abelian phases.  It has been argued that tunneling between the layers can transform the $(330)$-state into a phase supporting Fibonacci anyons \cite{vaezi14}.  These anyons are defined by simple fusion rules, but still allow for universal quantum computing \cite{nayak08}. Other non-Abelian phases are obtained via $p$-type pairing, either between particles within a layer or between all particles, leading to intra- and the inter-layer Pfaffian wave functions \cite{ardonne02,maissam10}. Recently, extensive numerical works have revealed some of these phases if interactions are properly modified \cite{Geraedts15,LiuVaezi15,Peterson15}. In particular, studies on the thin torus \cite{vaezi14} as well as exact numerics \cite{LiuVaezi15} point towards a Fibonacci phase if the short-range contribution to the interlayer interactions is weakened.

\begin{figure}
\includegraphics[width=8.7cm]{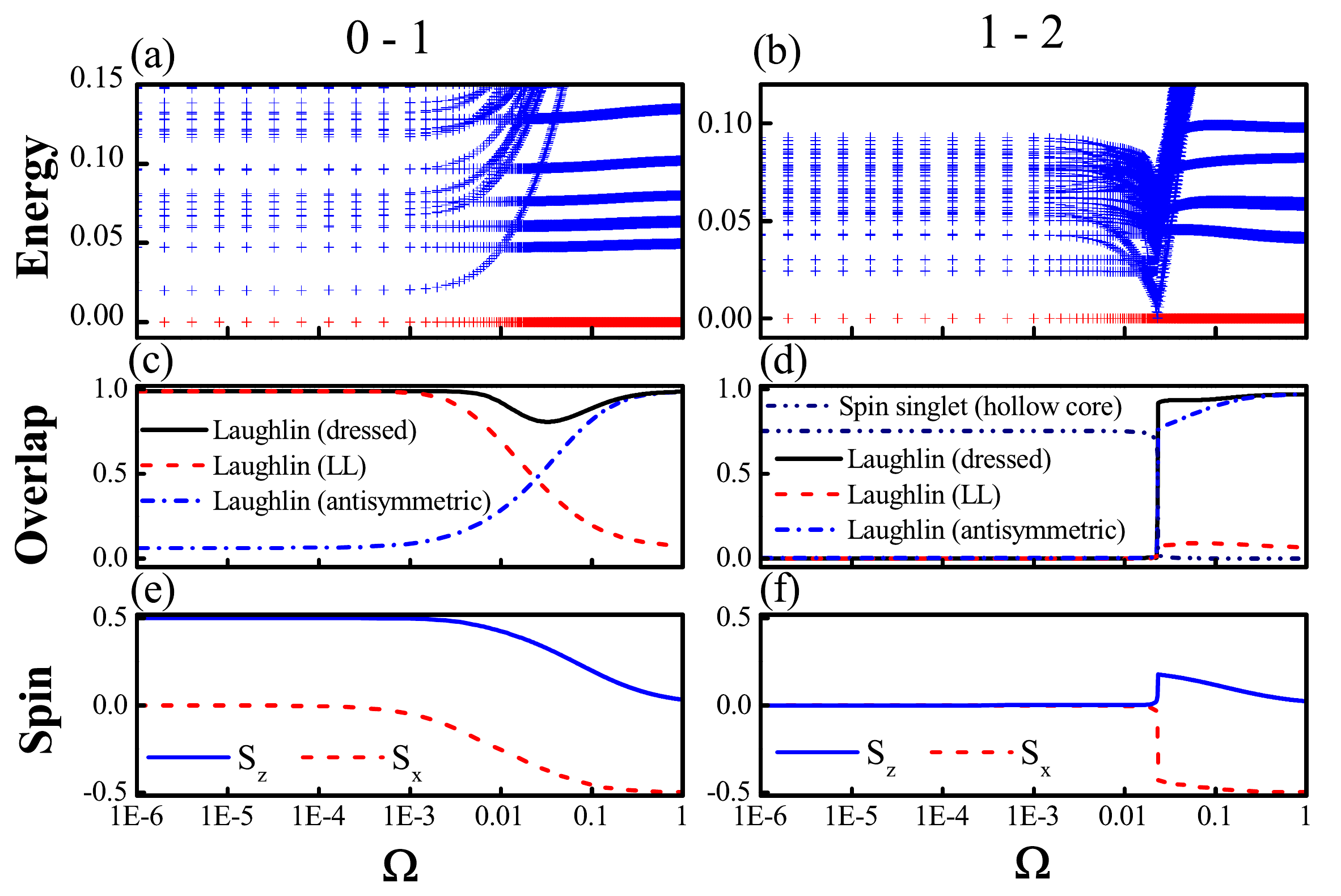}
\caption{\label{fig:EndepO} (a,b) Energy levels (above ground state in units of $e^2/\epsilon l^{}_B$) vs. Rabi frequency $\Omega$, for coupling ${\rm LL}_{0-1}$ (a), and ${\rm LL}_{1-2}$ (b). (c,d) Ground state overlaps with trial wave functions (particle-hole conjugate $1/3$ Laughlin state and a singlet phase obtained from hollow-core model). Trial states are constructed in three different bases: (1) \textit{LL basis.} All the electrons reside in the lower LL. (2) \textit{Dressed basis.} All electrons reside in lower eigenstates of Eq.~(\ref{eq:dress}), i.e. $|j\rangle \propto \left(\delta\-\sqrt{\delta^2+4\Omega^2}\right)|M+1,j\rangle+2\Omega |M,j\rangle$. (3) \textit{Antisymmetric basis.} All electrons reside in the singlet state, i.e.,  $|j\rangle \propto -|M+1,j\rangle+ |M,j\rangle$. 
(e,f) Spin polarization $S_\alpha = \frac{1}{2N} \sum_j \langle \sum_j \tau_\alpha^{(j)} \rangle$ of the ground state vs. $\Omega$ for ${\rm LL}_{0-1}$ (e), and ${\rm LL}_{1-2}$ (f).
Data in all panels (a--f) was obtained for 8 electrons on the torus, and $\delta=0.02$.
}\end{figure}

In both coupling scenarios, ${\rm LL}_{0-1}$ and ${\rm LL}_{1-2}$, ED on torus and sphere gives clear hints for a hole-conjugate Laughlin phase when the Rabi frequency is sufficiently strong. If the Laughlin state is formulated in a dressed LL basis, overlaps with this state reach close to 1, see  Fig.~\ref{fig:EndepO}(c,d). As already observed at $\nu=1/2$, the two coupling scenarios show different behavior when $\Omega$ is decreased. Again, while for ${\rm LL}_{0-1}$ tuning the Rabi frequency only rotates the spin, a transition into a singlet phase occurs for ${\rm LL}_{1-2}$, see Fig.~\ref{fig:EndepO}(e,f). In contrast to $\nu=1/2$, where the transitions occurs between two gapless phases, we now observe a transition between gapped phases, and the gap vanishes only at the critical point, see Fig.~\ref{fig:EndepO}(b). Also, at $\nu=2/3$, the transition does not affect the symmetry of the ground state (${\bf K}=(0,0)$ on both sides). 

The identification of the singlet phase at weak ${\rm LL}_{1-2}$ coupling is challenging. On the sphere, where our numerics extend up to 12 electrons, we find large gaps for $N=8$ and $N=12$, but tiny gaps for $N=6$ and $N=10$, suggesting a tetra-periodic system behavior. While an intralayer Pfaffian state, requiring ${\rm mod}(N,4)=0$, would explain this pattern, the overlap with this state is zero (for $N=8$ on sphere and disk). In contrast, significant overlaps are obtained with the interlayer Pfaffian state (see Table \ref{table}). However, the corresponding $(3q)$-fold torus degeneracy is not seen for 8 or 10 electrons. Lacking obvious ground state degeneracies beyond the $q$-fold center-of-mass degeneracy, an Abelian phase such as Jain's spin-singlet state seems possible \cite{JainBook,Belkhir93,Moran12}, but only infitesimal overlap is found. Given the relative weakness of $V_0^{\rm inter}$, we shall also consider the Fibonacci phase. On the torus, it is characterized by $2q$ ground states at ${\bf K}=(0,0)$ \cite{LiuVaezi15}. While we obtain the second and the third state at ${\bf K}=(0,2)$ and ${\bf K}=(2,0)$ on an isotropic torus, squeezing the torus changes this pattern, and the lowest two eigenstates indeed become singlets at ${\bf K}=(0,0)$. Moreover, they have large overlaps with the corresponding eigenstates of the hollow-core Hamiltonian (0.76 and 0.81 on an isotropic torus), previously identified as representatives of the Fibonacci phase \cite{LiuVaezi15}. This makes the Fibonacci phase more likely than other candidate phases, although a final conclusion is impossible based on the available numerical results.

\textit{Thermalization.} In this work, we have assumed that the electronic system thermalizes to the ground state in the rotating frame of the optical drive field. To estimate the validity of this approximation, we must compare the timescale for relaxation of the optically excited Landau levels to the timescale for thermalization of the electronic system with the lattice.  The carrier lifetime of optically excited Landau levels has contributions from optical relaxation, phonon relaxation, and Auger scattering into other Landau levels \cite{Funk15}.  In Ref.~\cite{Mittendorff14}, it was measured at moderate magnetic fields in epitaxial graphene samples to be roughly 10 - 20 ps.  Although one expects longer lifetimes in higher quality graphene samples suitable to observe the FQH effect, we can use this as an upper bound on the relaxation rate.  In units $e^2/\epsilon l^{}_B$, this timescale translates to roughly $10^{-3}$ to $10^{-4}$, depending on the magnetic field.  For ${\rm LL}_{0-1}$ coupling, the  Laughlin state of the driven and the non-driven regime are adiabatically connected, and one can adiabatically prepare the system by slowly turning on the light. In contrast, the singlet states for ${\rm LL}_{1-2}$ coupling cannot be connected to the non-driven regime, which makes the thermalization problem particularly relevant.  For the case of the $\nu=2/3$ singlet phase, we can roughly estimate the thermalization time by the size of the many-body gap in the spectrum, which, from Fig.~\ref{fig:EndepO}, is on the order of $10^{-2}$.   As a result, there is a large separation of timescales between the thermalization and carrier relaxation, which allows the system to remain in the rotating frame ground states before carrier relaxation.  For the gapless phases at $\nu=1/2$ the system will still thermalize in the rotating frame, however, the timescale is more difficult to estimate as it depends on the slowest diffusive modes in the system.  
%Another important consideration is the heating of the sample due to thermalization with lattice phonons.  
A more detailed study of the thermalization dynamics in this regime is beyond the scope of the present work; however, it is worth noting that there has been recent progress in understanding of thermalization of driven isolated systems \cite{Lazarides:2014cl,DAlessio:2014fg,Ponte:2015hm} and also thermalization of Floquet systems coupled to a bath \cite{Dehghani:2014ut,Seetharam:2015ib}. It has been pointed out that electron-phonon interaction and specific Fermi reservoirs could lead to thermalization of the system in the rotating frame, in the long time steady-state limit \cite{Seetharam:2015ib}.

In conclusion, we have considered single layer graphene in the FQH regime with an optical field in resonance with a LL transition. The proposed scheme synthesizes a two-component FQH system, with the light field playing the role of tunneling between two layers. For weak tunneling between ${\rm LL}_1$ and ${\rm LL}_2$, a many-body singlet phase is formed at $\nu=1/2$ and $\nu=2/3$. In contrast, strong tunneling and/or tunneling between ${\rm LL}_0$ and ${\rm LL}_1$ lead to a polarized phase within the lower dressed LL. Our study gives new impetus towards experimental realization of multicomponent FQH states and \textit{in situ} control of the phase transition using externally applied optical fields and graphene. A similar scheme could also be applied to other 2D materials with Dirac bands, such as monolayer transition metal dichalcogenides \cite{Xiao12,Wang16}.

\acknowledgments
\textit{Acknowledgments.} 
We acknowledge fruitful comments by S. Simon, E. Demler, K. Seetharam, and S. Das Sarma. This research was supported under National Science Foundation Grants EFRI-1542863, PFC at the Joint Quantum Institute, CNS-0958379, CNS-0855217, ACI-1126113 and the City University of New York High Performance Computing Center at the College of Staten Island, and AFOSR-MURI  FA95501610323, Sloan Fellowship, YIP-ONR.

\putbib
\end{bibunit}

\begin{bibunit}
	
\clearpage
\onecolumngrid
\setcounter{figure}{0}
\makeatletter
\renewcommand{\thefigure}{S\@arabic\c@figure}
\setcounter{equation}{0} \makeatletter
\renewcommand \theequation{S\@arabic\c@equation}
\renewcommand \thetable{S\@arabic\c@table}

\begin{center} 
	{\large \bf Supplementary online material for the paper
		\protect \\{\it \bf Light-induced fractional quantum Hall phases in graphene}}
\end{center}

\section{Exact diagonalization technique and supplemental numerical data}
\subsection{Disk}
\subsubsection{Basis states}
On a disk the perpendicular magnetic field is most conveniently taken into account by the symmetric gauge ${\bf A} \sim (y,-x,0)$. The eigenstates of $({\bf p}-e{\bf A})^2/2,$ are then described by the LL index $n$ and a quantum number which is related to the $z$-component of angular momentum $\ell_z=\hbar(m-n)$. The (unnormalized) single-particle wave functions read 
\begin{align}
	\phi_{m,n}(z) \sim \exp\left[|z|/(2l_{\rm B})^2\right] z^{m-n} L_n^{m-n}\left[|z|/(2l_{\rm B}^2)\right],
\end{align}
with $L_n^m(x)$ denoting the generalized Laguerre polynomials. The graphene eigenstates are spinors composed by wavefunctions  $\phi_{m,n-1}(z)$ and $\phi_{m,n}(z)$, as described in the main text.

Although a real quantum Hall sample usually is flat, as is the disk geometry, calculations on the disk might fail to describe the physics in the real system due to finite-size effects.
In contrast to curved geometries (torus, sphere), the disk is not compact and thus it exhibits both bulk and edge states. Moreover, there is no natural cutoff for the $m$ quantum number, and thus the filling factor (i.e. electrons per states in a given LL) cannot be defined. Instead, the state of a few- or many-body system  with $N$ electrons is characterized by $M=\sum_i m_i$, which in the lowest LL coincides with the $z$-component of total angular momentum. This allows for performing exact diagonalization in finite Hilbert spaces characterized by $N$ and $M$, and provides a strict cutoff for single-particle momenta $m_i\leq M-(N^2-3N+2)/2$. In practice, though, single-particle momenta $m_i$ relevant for low-energy physics are much smaller, and the cutoff can be chosen differently.

\subsubsection{Yrast line}
\begin{figure*}
	\includegraphics[width=0.5\textwidth]{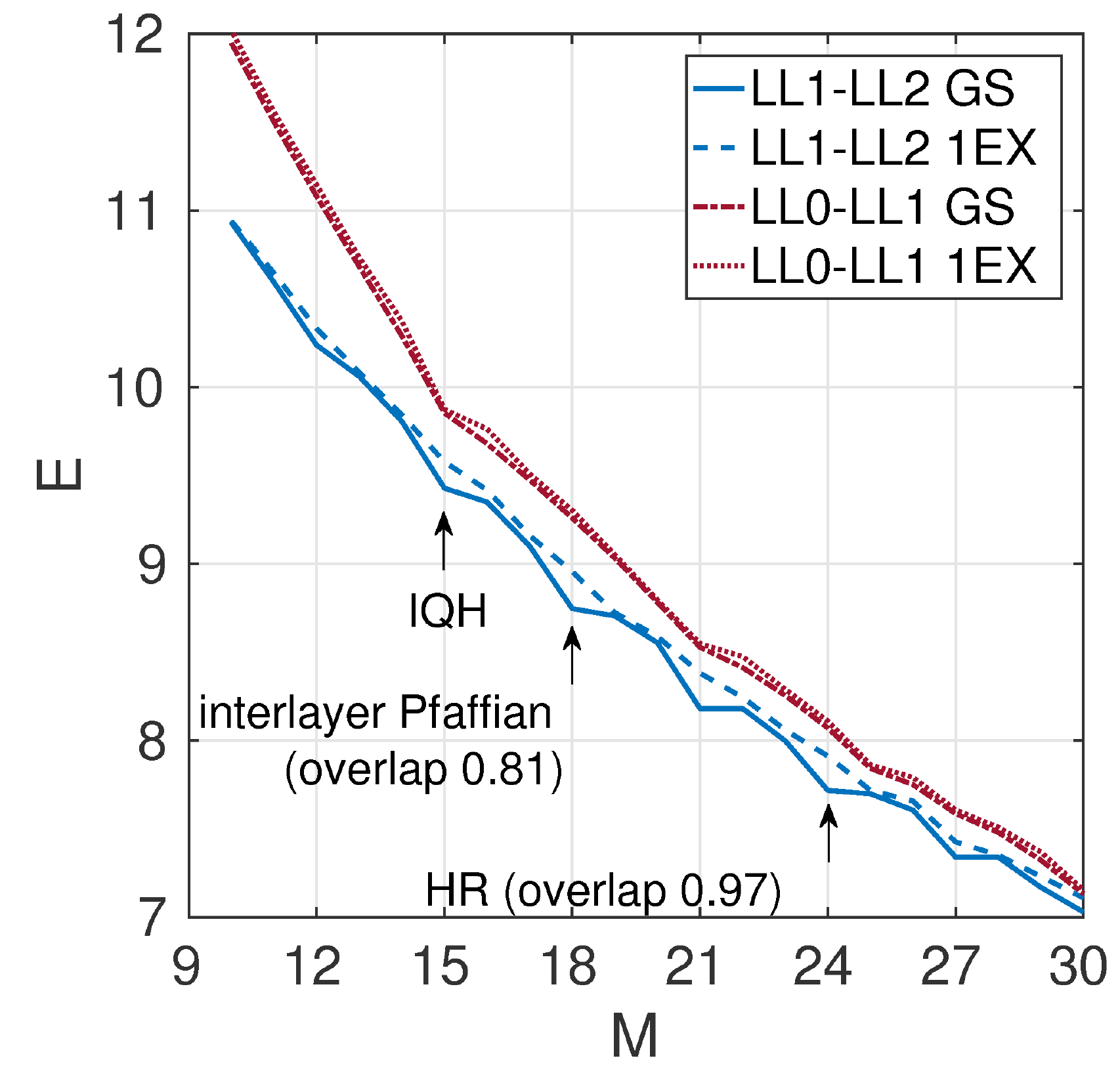}
	\caption{\label{fig:yrast} Energy of ground state and first excited state, for $N=6$ electrons on a disk, as a function of $M$. We assume a coupling of ${\rm LL}_0$ and ${\rm LL}_1$ (red lines), or ${\rm LL}_1$ and ${\rm LL}_2$ (blue lines), at detuning $\delta=0.02$, and $\Omega=10^{-3}$. 
		Pronounced downward cusps, together with large gaps, are found for the coupling between ${\rm LL}_1$ and ${\rm LL}_2$ at $M=15,18,21,24,27$, giving a hint for incompressible phases.
		Units of energy are $e^2/(\epsilon l_{\rm B})$.}
\end{figure*}

In order to find hints for the phases which might be exhibited also in a larger system, we have scanned, at fixed particle number $N$, a range of values $M$. The behavior of the ground state energy as a function of $M$, the so-called Yrast line, is shown in Fig.~\ref{fig:yrast}, for couplings ${\rm LL}_{0-1}$ and ${\rm LL}_{1-2}$. At certain values of $M$, downward cusps in $E(M)$ give hints for incompressible phases, since decreasing $M$ from these values is energetically costly, and $M$ also parametrizes the system size. Notably, these cusps are much more pronounced for coupling ${\rm LL}_{1-2}$. In this case, all cusps seen in 
Fig.~\ref{fig:yrast} (for $N=6$ electrons, cusps at $M=15,18,21,24,27$) come along with a relatively large gap, and the ground states are fully unpolarized, i.e. $P\equiv (N_\uparrow-N_\downarrow)/N=0$. In contrast, for ${\rm LL}_{0-1}$, all downward cusps (at $M=15,21,25$) stem from almost fully polarized grond states, i.e. $P\approx N$. This observation confirms our expectation that coupling ${\rm LL}_{1-2}$ supports the formation of singlets, while coupling ${\rm LL}_{0-1}$ does not.

\subsubsection{Overlaps}
Further, we have tried to identify the ground states at the cusps by determining the overlap with trial wave functions. On the disk geometry, decomposing a given $N$-body wave function into Fock basis states can straightforwardly be achieved for small systems (see Ref. \cite{strongdeco}). We then denote as ``overlap'' the scalar product between this Fock representation of the trial wave function, and the Fock representation of our numerically obtained ground state. Of course, this number would only correspond to the spatial overlap if the Fock basis sets are the same. Here, however, the numerical wave functions are obtained in different graphene LLs, while trial wave functions are defined in the lowest non-relativistic LL (i.e. $\phi_{m,0}(z)$). Yet, since a one-to-one mapping between the two Hilbert spaces exists, our measure of overlap still serves to compare the correlations in both states.

The ground state at the cusp at $M=18$ has significant overlap (0.81) with the interlayer Pfaffian wave function. A smaller overlap of 0.68 is found with the 330-Halperin state, i.e. a combination of two independent 1/3-Laughlin states. The overlap with the intralayer Pfaffian state is zero. Unfortunately, we are unaware of a wave function describing the Fibonacci phase which we expect to compete with these phases. We have also considered a system of $N=8$ electrons at $M=36$, where the overlap with the interlayer Pfaffian state still reaches 0.63.

At $M=24$, the Haldane-Rezayi wave function can be constructed (which in the thermodynamic limit corresponds to $\nu=1/2$), but also a composite-fermion state corresponding to $\nu=2/3$. The latter is obtained by multiplying the flux attaching Jastrow factor $\prod_{i<j}(z_i-z_j)^2$ with Slater determinants for the composite fermions \cite{wu12}. For six particles, the Slater determinant has $M=30$, so the target value $M=24$ can be reached by filling the three composite fermions of each component into flux-reversed LLs, with $m_i=0,-1,-2$. Evaluation of the overlaps shows a clear favor for the Haldane-Rezayi phase (overlap 0.97), while the overlap with the composite fermion wave function is almost zero. 

At $M=27$, the Haperin $331$-state can be constructed, but its overlap with the ground state is infinitesimal. We are not aware of any unpolarized trial wave functions which might correspond to the cusp at $M=21$. In summary, overlap calculations on the disk show good overlap with the Haldane-Rezayi phase corresponding to half-filling, and average overlap with the interlayer Pfaffian phase at 2/3-filling. However, there are hints for other incompressible phases without known trial wave function description. Therefore, we cannot exclude the possibility that these states correspond to the same filling factors, which would give rise to a competition between different phases in the thermodynamic limit.

\subsection{Torus}
The advantage of torus geometry relies in the fact that it allows to obtain topological degeneracies of the ground state. In this case we work in Landau gauge, namely $A=B(-y,0,0)$. We take the torus to be rectangular, characterized by the dimensions $L_x$ and $L_y$ ($\lambda=L_x/L_y$ defining axis ratio). After imposing periodic boundary conditions (PBC), the single particle wave functions are \cite{ChakrabortyBook}
\begin{equation}
	\phi_{n,j}(x,y)=C^{}_n\sum_{m=-\infty}^{\infty}e^{\frac{i}{l^2_B}\left(Y^{}_j-mL^{}_y\right)x} e^{-\frac{\left(y+mL^{}_y-Y^{}_j\right)^2}{2l^2_B}}H^{}_n\left(\frac{y+mL^{}_y-Y^{}_j}{l^{}_B}\right),
\end{equation}
where $C^{}_n=\left(L^{}_x\sqrt{\pi}l^{}_B2^nn!\right)^{-\frac12}$, $Y^{}_j=kl^2_B=2\pi jl^2_B/L^{}_x$. Here $j$ is an integer, which takes the values $0,1\cdots N^{}_s-1$, $N^{}_s=L^{}_xL^{}_y/2\pi l^2_B$ describing the degeneracy of each LL and corresponds to the number of elementary magnetic flux threading the torus. $j$ is the quantum number characterizing momentum in $x$ direction, which is a conserved quantity due to the translational invariance. For the many-body problem the filling factor is defined as $\nu=N^{}_e/N^{}_s$. Using the single particle eigenstates of graphene the interaction matrix element of (3) in main text can be written in the form \cite{ChakrabortyBook}
\begin{equation}
	A^{n_1,j_1,n_2,j_2}_{n_3,j_3,n_4,j_4}=\frac{\delta^\prime_{j_1+j_2,j_3+j_4}}{2L^{}_xL^{}_y}\sideset{}{'}\sum_\mathbf{q}\delta^\prime_{j_1-j_4,q^{}_xL^{}_x/2\pi}V(q)e^{iq^{}_y\left(Y^{}_{j_1}-Y^{}_{j_3}\right)}\mathcal{F}_{n_1,n_4}(\mathbf{q})\mathcal{F}_{n_2,n_3}(-\mathbf{q}),
\end{equation}
where primed summation excludes $\mathbf{q}=0$ term and primed Kronecker $\delta^\prime$ is defined $\mathrm{mod}\,N_s$. $V(q)$ is the Fourier transform of the Coulomb interaction, and $\mathcal{F}_{n_1,n_2}(\mathbf{q})$ are the form factors of the Landau levels. The graphene form factor are given by \cite{Goerbig11}
\begin{equation}
	\mathcal{F}_{n_1,n_2}(\mathbf{q})=e^{-\frac{|q|^2l^2_B}{4}}\sqrt{\frac{(n_2-1)!}{(n_1-1)!}}\left(-\frac{\bar{q}l_B}{\sqrt{2}}\right)^{n_1-n_2}\left[C_{n_1}^{-}C_{n_2}^{-}L_{n_2-1}^{n_1-n_2}\left(\frac{|q|^2l^2_B}{2}\right)+C_{n_1}^{+}C_{n_2}^{+}\sqrt{\frac{n_2}{n_1}}L_{n_2}^{n_1-n_2}\left(\frac{|q|^2l^2_B}{2}\right)\right],
	\label{FormFactor1}
\end{equation}
when $n_1\geq n_2$ and
\begin{equation}
	\mathcal{F}_{n_1,n_2}(\mathbf{q})=e^{-\frac{|q|^2l^2_B}{4}}\sqrt{\frac{(n_1-1)!}{(n_2-1)!}}\left(\frac{ql_B}{\sqrt{2}}\right)^{n_2-n_1}\left[C_{n_1}^{-}C_{n_2}^{-}L_{n_1-1}^{n_2-n_1}\left(\frac{|q|^2l^2_B}{2}\right)+C_{n_1}^{+}C_{n_2}^{+}\sqrt{\frac{n_1}{n_2}}L_{n_1}^{n_2-n_1}\left(\frac{|q|^2l^2_B}{2}\right)\right],
	\label{FormFactor2}
\end{equation}
when $n_1<n_2$ and we use the notation $q=q_x+iq_y$ and $\bar{q}=q_x-iq_y$ in this section. 
%Defining as $V^{n_1n_2n_3n_4}_\mathrm{eff}=V(q)\mathcal{F}_{n_1,n_4}(\mathbf{q})\mathcal{F}_{n_2,n_3}(-\mathbf{q})$ and doing the expansion
%\begin{equation}
%V^{n_1n_2n_3n_4}_\mathrm{eff}=4\pi l^2_B\sum_{m}V^{n_1n_2n_3n_4}_{m}L^{}_m\left(q^2l^2_B\right)e^{-\frac{q^2l^2_B}{2}},
%\label{Expansion}
%\end{equation}
%the matrix elements can be written in an alternative form
%\cite{Jeong15}
%\begin{equation}
%\label{AltForm}
%A^{n_1,j_1,n_2,j_2}_{n_3,j_3,n_4,j_4}=\delta^\prime_{j_1+j_2,j_3+j_4}\sum_{m=0}^{\infty}\frac{V_{m}^{n_1n_2n_3n_4}}{N_s} 
%	\times\sideset{}{'}\sum_{\mathbf{q}}\delta^\prime_{j_1-j_4,q_xL_x/2\pi}e^{iq_y(Y_{j_1}-Y_{j_3})}e^{-\frac{q^2l_{B}^2}{2}}L_m(q^2l_B^2).
%\end{equation}
%In (\ref{Expansion}) and (\ref{AltForm}) $V_{m}^{n_1n_2n_3n_4}$ are the Haldane pseudopotentials \cite{PrangeBook}, which can be obtained from (\ref{Expansion}) and orthogonality of Laguerre polynomials
%\begin{equation}
%V_{m}^{n_1,n_2,n_3,n_4}=\frac{1}{4\pi^2}\int d^2qV(q)e^{\frac{-q^2l_B^2}{2}}L_m(q^2l_B^2)\mathcal{F}_{n_1,n_4}(\mathbf{q})\mathcal{F}_{n_2,n_3}(-\mathbf{q}).
%\label{HaldPseudo}
%\end{equation}
%The form of the matrix elements (\ref{AltForm}) can also be used to obtain trial wave function with parent Hamiltonian characterized with specific non-zero pseudopotentials. 

As has been shown before \cite{Haldane85,ChakrabortyBook} for many-body system there is a relative translation operator $T^i_\mathrm{R}\left(p\mathbf{L}_{mn}\right)$ which commutes with the full Hamiltonian. Here $p$ is defined by filling factor $\nu=p/q$, where $p$ and $q$ are coprime integers and $\mathbf{L}_{mn}=mL_x\hat{\mathbf{x}}+nL_y\hat{\mathbf{y}}$ is the translation lattice vector. The relative translation operator acts on many-body state $|n_1j_1,n_2j_2\dots, n_{N_e}j_{N_e}\rangle$
\begin{equation}
	T^i_\mathrm{R}\left(p\mathbf{L}_{mn}\right)|n_1j_1,n_2j_2\dots, n_{N_e}j_{N_e}\rangle=(-1)^{pqmn\left(N_e+1\right)}e^{-i\frac{2\pi mK_x}{N}}|n_1(j_1+nq),n_2(j_2+nq)\dots, n_{N_e}(j_{N_e}+nq)\rangle,
\end{equation} 
where $N_e=pN$ and $K_x=\sum_{i}^{N_e}j_i\mathrm{mod}N$ is the total momentum in $x$ direction. We use the quantum numbers $K_x$ and $K_y$ characterizing the eigenvalues of $T^i_\mathrm{R}\left(p\mathbf{L}_{mn}\right)$ to classify the eigenstates of the Hamiltonian. 

\begin{figure*}
\includegraphics[width=0.99\textwidth]{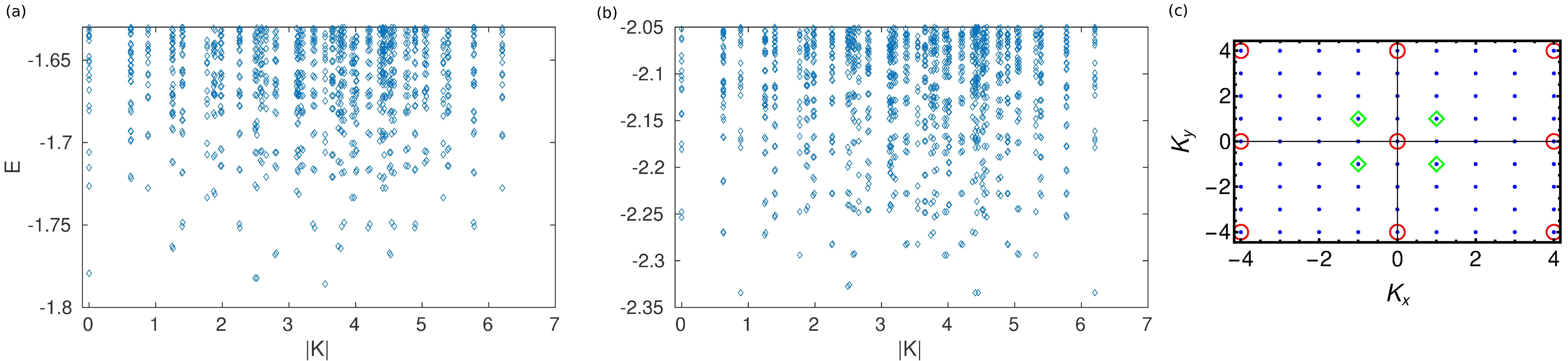}
\caption{\label{fig:Kspec12} (a,b) Energy spectra of 8 electrons with 16 fluxes ($\nu=1/2$) on the torus (axis ratio 0.99), when ${\rm LL}_1$ and ${\rm LL}_2$ are weakly coupled with Rabi frequency $\Omega=0.01$ (a), or strongly coupled with Rabi frequency $\Omega=0.1$ (b), at detuning $\delta=0.02$. All energies are in units  $e^2/\epsilon l_{\rm B}$. (c) First Brillouin zone of the Haldane pseudomomenta, with the red squares marking the 4 quasi-degenerate minima in the weak-coupling phase, and the green diamonds marking the 4 fully degenerate minima in the strong-coupling phase. The 4 ground states in the weak-coupling phase have significant overlap with the HR state (zero-energy eigenstates of hollow-core potential), namely 0.83 at $|K|=0$, and 0.72 at other values of $K$.
}\end{figure*}
	
\begin{figure*}
\includegraphics[width=0.99\textwidth]{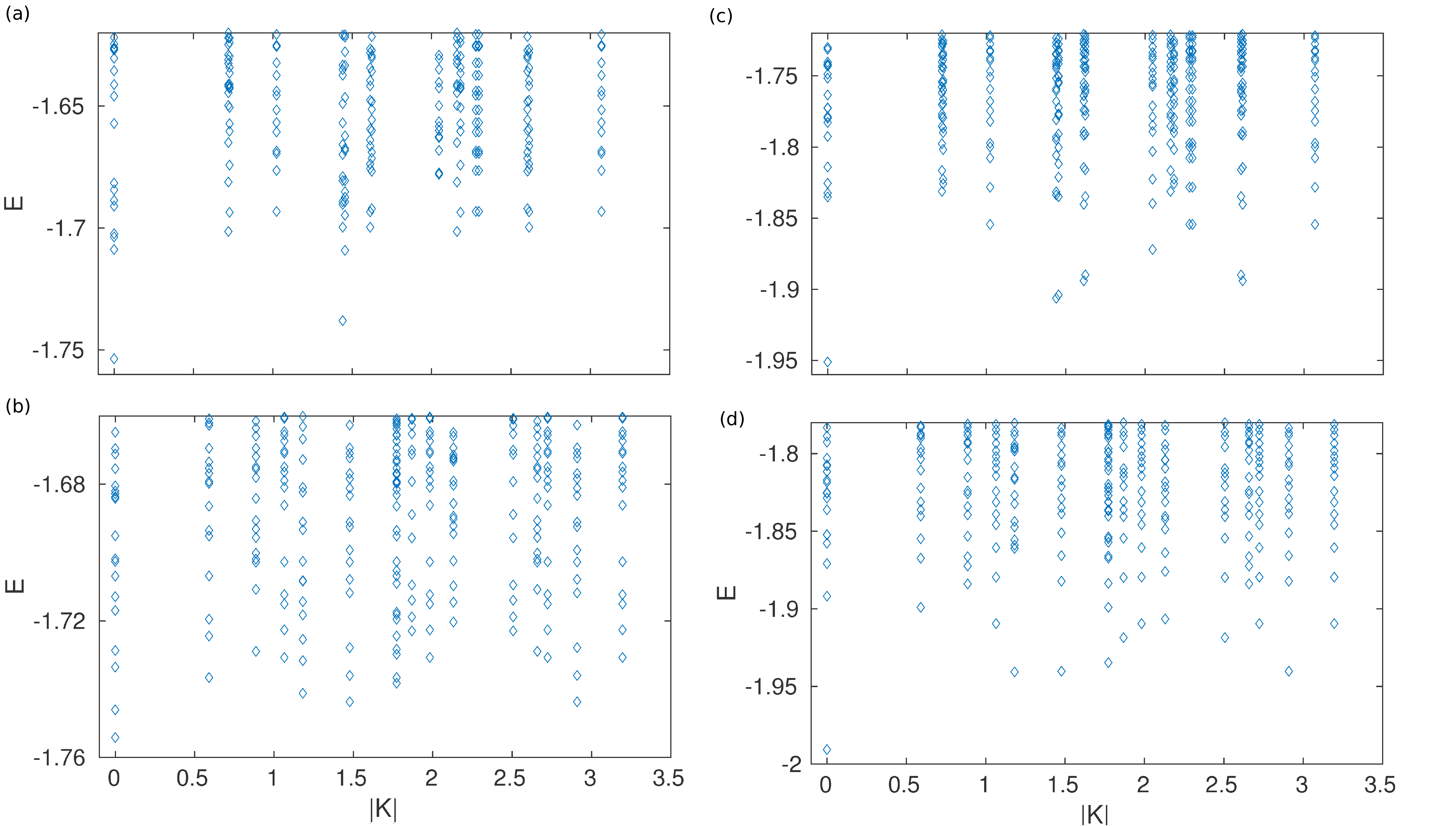}
\caption{\label{fig:Kspec123} (a--d) Energy spectra of 8 electrons with 12 fluxes ($\nu=2/3$) on the torus at axis ratio 0.99 (a,c), and axis ratio 1.5 (b,d). The coupling is between ${\rm LL}_1$ and ${\rm LL}_2$,  at weak Rabi frequency $\Omega=10^{-3}$ (a,b), and at strong Rabi frequency $\Omega=0.05$ (c,d), with detuning $\delta=0.02$. All energies are in units  $e^2/\epsilon l_{\rm B}$. While there is a unique ground state at $|K|=0$ in the strong coupling regime independently from the axis ratio, the pattern in the weak coupling regime is less evident. A two-fold quasidegeneracy, at $|K|=0$ and $|K|\approx\pi/2$, is seen for axis ratio 0.99, but squeezing the torus renders two states at $|K|=0$ to be the ones of lowest energy. Such degeneracy would be compatible with a Fibonacci phase. 
}\end{figure*}

In Fig.~\ref{fig:Kspec12} the spectra for 8 electron at filling factor $\nu=1/2$ is shown obtained on the torus with axis ratio 0.99, when $\mathrm{LL}_1$ and $\mathrm{LL}_2$ are weakly coupled with Rabi frequency $\Omega=0.01$ (a) and strongly coupled with Rabi frequency $\Omega=0.1$ (b). At weakly coupled case the four ground states are located at the center and edges of the Brillouin zone, are spin-singlet states and have considerable overlap with Haldane-Rezayi state. At strongly coupled case the location of four ground states in the Brilloin zone changes and the system is spin-polarized, indicating compressible Fermi liquid phase. Similar spectra for filling factor $\nu=2/3$ is shown in Fig.~\ref{fig:Kspec123} for two different axis ratios 0.99 (a,c) and 1.5 (b,d) for both weak and strong coupling cases. In weak coupling case the ground state at $|K|=0$ is again spin-singlet although it is only triple degenerate due to the center of mass degeneracy. Squeezing the torus makes two lowest states to be located at $|K|=0$, which denotes six-fold degeneracy and is compatible with Fibonacci phase \cite{vaezi14,LiuVaezi15}. In strong coupling phase the ground state is unique regardless of torus axis ratio and is spin-polarized indicative of particle-hole conjugate Laughlin 2/3 phase.  
	
\subsection{Sphere}
The exact-diagonalization calculations on a sphere has a similar structure as on the disk, although the sphere being compact does not suffer from mixing of the bulk and edge states. We use the calculations on the sphere mostly for obtaining overlaps with proposed trial states. The magnetic field on a sphere is produced by placing magnetic monopole with the strength $Q$ at the center of the sphere \cite{Haldane83,Fano86} and this produces $N_s=2Q$ elementary magnetic flux threading the sphere. Due to the spherical symmetry both total angular momentum and its projection are conserved quantities. The single particle eigenstates in the lowest LL are obtained for angular momentum $l=Q$ and have the form
\begin{equation}
\phi_{Qm}(u,v)=\left[\frac{2Q+1}{4\pi}\left(\begin{array} {c}
	2Q \\ Q-m
\end{array}\right)\right]^\frac{1}{2}(-1)^{Q-m}u^{Q+m}v^{Q-m}\cdot,
\end{equation}
where $u=\cos\left(\theta/2\right)e^{i\phi/2}$ and $v=\sin\left(\theta/2\right)e^{-i\phi/2}$ are the spinor variables and quantum number for angular momentum in $z$ direction $m$ takes the values $-Q,-Q+1\dots, Q$. Therefore, the degeneracy of the LL in this case is $2Q+1$. For many-body system the filling factor is related to number of electrons by $N_s=N_e/\nu-\mathcal{S}$, where $\mathcal{S}$ denotes the shift. For the calculation on the sphere, different LLs are considered as layer indices all described by lowest LL eigenstates. The structure of different LLs is completely encoded in Haldane pseudopotentials, which define the interaction Hamiltonian
\begin{equation}
A^{n_1,m_1,n_2,m_2}_{n_3,m_3,n_4,m_4}=\sum_{L=0}^{2Q}\sum_{M=-L}^{L}\langle Qm_1,Qm_2|LM\rangle V^{n_1n_2n_3n_4}_{2Q-L}\langle LM|Qm_3,Qm_4\rangle,
\end{equation}
where $\langle LM|Qm_1,Qm_2\rangle$ are the Clebsch-Gordon coefficients, $L$ and $M$ are the total angular momentum and $z$ projection of it for the pair and $V^{n_1n_2n_3n_4}_{m}$ are the Haldane pseudopotentials. For Coulomb interaction we use the values of the pseudopotentials obtained on the plane, which should be a good approximation in thermodynamic limit. This form of interaction potential can be used also to obtain trial wave functions described by two-body interaction and specific non-zero pseudopotentials. Besides that three-body interaction Hamiltonian can be straightforwardly defined on the sphere as well, which can be used to obtain trial wave function for some non-Abelian phases, such as Moore-Read \cite{RR96} and Interlayer Pfaffian state \cite{Lankvelt04,Ardonne11}. 

\begin{figure*}
\includegraphics[width=0.8\textwidth]{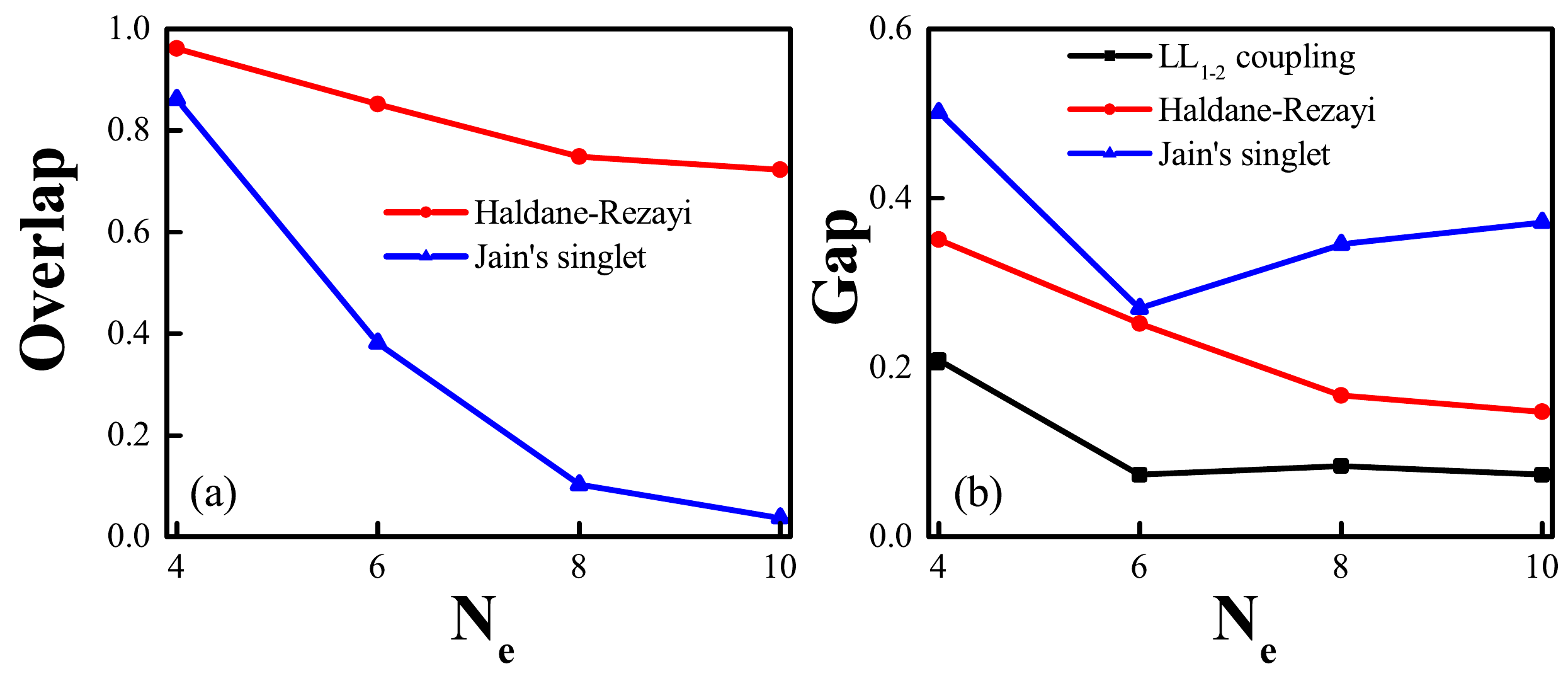}
\caption{\label{fig:HRJainComp} 
	(a) The dependence of the overlap of the ground state for coupling $\mathrm{LL}_{12}$ with Haldane-Rezayi and Jain's singlet states on system size. (b) The dependence of the gap for the case of coupling $\mathrm{LL}_{12}$, HR and Jain's singlet phases on the system size. The calculations are done on the sphere for filling factor $\nu=1/2$ at detuning $\delta=0.02$, and $\Omega=10^{-3}$. The HR and Jain's singlet states are obtained with non-zero pseudopotentials equal to one. The gap for the case of coupling $\mathrm{LL}_{12}$ is normalized by the value of $V^\mathrm{inter}_1$ pseudopotential.}
\end{figure*}

For $\nu=1/2$ at weak light-matter coupling and for spin singlet phase the possible trial states are the (331) Halperin state, HR state and Jain's singlet \cite{Belkhir93,Moran12}. The (331) state is described by the shift $\mathcal{S}=3$. For that shift the ground state of the system considered in this work correspond to the state with total angular momentum $L=2$ for 8 electrons. This shows that ground state is not translationally invariant phase at that shift and (331) is not a good candidate. The HR and Jain's singlet state are characterized with the shift $\mathcal{S}=4$. HR is a gapless phase, whereas Jain's singlet is gapped. HR state is the exact zero energy eigenstate of hollow core Hamiltonian (only $V^\mathrm{intra}_{1}$ and $V^\mathrm{inter}_{1}$ being non-zero). Since constructing the exact wavefunction of a Jain’s singlet state for $\nu=1/2$ is	numerically expensive, instead, we use the ground state of a Hamiltonian with pseudopotentials  $V_{0}=V_{1}$, which is known to have a large overlap with the Jain’s singlet state [14]. In Fig.~\ref{fig:HRJainComp} (a) we compare overlaps of the ground state of the singlet phase at Rabi frequency $\Omega=10^{-3}$ for the system considered in this work with HR state and the Jain's singlet state. As can be seen from the figure the overlap with HR state is always larger and it drops much slower with system size than Jain's singlet. In Fig.~\ref{fig:HRJainComp} (b) the dependence of the gap of these three systems on system size is presented. The HR state and Jain's singlet state are obtained with non-zero pseudopotentials equal to one and the gap for singlet state at coupling $\mathrm{LL}_{12}$ is normalized by the value of $V^\mathrm{inter}_1$. As can be seen from the figure the gap of HR state decreases with system size, whereas the gap of Jain's singlet fluctuates. While the gap for singlet phase for coupling $\mathrm{LL}_{12}$ does not show clear trend of decreasing with system size, it is small compared to other systems. This non-decreasing of the gap with system size is possibly related to the small system sizes considered in the calculation and this leads us conclude that the spin singlet phase observed at $\nu=1/2$ is HR phase.         

For $\nu=2/3$ filling factor and spin-singlet phase we have considered the following trial states for overlap calculation: Jain's composite fermion phase ($\mathcal{S}=1$), (330) phase ($\mathcal{S}=3$), Interlayer and Intralayer Pfaffian phases ($\mathcal{S}=3$). We get almost zero overlap with Jain's composite fermion and Intralayer Pfaffian phases. There is a sizable overlap with (330) and Interlayer Pfaffian state, although the overlap with Interlayer Pfaffian is always bigger. We get at Rabi frequency $\Omega=10^{-3}$ 0.99 overlap for $N_e=4$, 0.55 for $N_e=8$ and 0.39 for $N_e=12$ with Interlayer Pfaffian phase. Therefore, the singlet state for $\nu=2/3$ is either Interlayer Pfaffian or Fibonacci phase, as was pointed out by torus calculation. Currently, it is unclear how to obtain Fibonacci trial state on the sphere, although the ground state of hollow core interaction gives larger overlap than Interlayer Pfaffian state (0.62 for $N_e=8$) and it is claimed that for $\nu=2/3$ hollow-core interaction supports Fibonacci phase as the ground state \cite{LiuVaezi15}.     

\section{Haldane pseudopotentials}
In this section we review in more detail the concept of Haldane pseudopotentials, and derive the appropriate pseudopotentials for our system.
Haldane pseudopotentials are derived from a series expansion for the matrix interactions elements, and typically only a few parameters of this expansion are enough to capture the physics. Therefore, interaction matrix elements are transformed from a two-particle basis $\ket{m_1,m_2;n_1,n_2}$, characterized by the individual quantum numbers $n_1$ and $n_2$ (LL indices), as well as  $m_1$ and $m_2$ (angular momenta), to a two-particle basis $\ket{m,M;n_1,n_2}$, where (in the lowest LL) $m$ and $M$ correspond to relative angular momentum and center-of-mass angular momentum \footnote{To generalize to arbitrary LLs, one can first define the basis in the lowest LL, and then apply LL raising operators.}. The matrix interaction elements transform according to:
\begin{align}
\bra{m_1,m_2;n_1,n_2}\hat V \ket{m_3,m_4;n_3,n_4} = \sum_{m,M,m',M'} \bra{m,M;n_1,n_2}\hat V \ket{m',M';n_3,n_4}.
\end{align}
As interactions do not depend on the center-of-mass motion, and a rotationally symmetric interaction will also not modify relative angular momentum, this can be re-written as
\begin{align}
\bra{m_1,m_2;n_1,n_2}\hat V \ket{m_3,m_4;n_3,n_4} = \sum_{m,M} V_m^{n_1,n_2,n_3,n_4} \braket{m_1,m_2;n_1,n_2}{m,M;n_1,n_2} \braket{m,M;n_3,n_4}{m_3,m_4;n_3,n_4}
\end{align}
with the Haldane pseudopotential defined as
\begin{align}
V_m^{n_1,n_2,n_3,n_4} = \bra{m,M;n_1,n_2}\hat V \ket{m,M;n_3,n_4},
\end{align}
which does not depend on $M$. For interactions which decay sufficiently fast with distance, contributions at large relative angular momentum $m$ will be small. So only a few pseudopotentials will be needed for calculations on a disk. Also note that for indistinguishable fermions (i.e. $n_1=n_2=n_3=n_4$) the relative angular momentum can only take odd values.

Evaluating the Haldane pseudopotentials leads to \cite{PrangeBook}
\begin{equation}
V_{m}^{n_1,n_2,n_3,n_4}=\frac{1}{4\pi^2}\int d^2qV(q)e^{\frac{-q^2l_B^2}{2}}L_m(q^2l_B^2)\mathcal{F}_{n_1,n_4}(\mathbf{q})\mathcal{F}_{n_2,n_3}(-\mathbf{q}),
\end{equation}
where $\mathcal{F}_{n_1,n_2}(\mathbf{q})$ form factors are defined in (\ref*{FormFactor1}) and (\ref*{FormFactor2}).

In our case, only two LLs will be relevant, so we will in the following replace the LL indices $n_i$ by symbols $\uparrow$ and $\downarrow$. After transforming Coulomb interactions into the frame rotating with the LL coupling, we have obtained a Kronecker $\delta_{n_1+n_2,n_3+n_4}$, so only the following Haldane pseudopotentials will be present:
intra-level pseudopotentials $V_m^\uparrow \equiv V_m^{\uparrow\uparrow,\uparrow\uparrow}$ and $ V_m^\downarrow = V_m^{\downarrow\downarrow,\downarrow\downarrow}$, as well as inter-level pseudopotentials $V_m^{\|} \equiv V_m^{\uparrow\downarrow,\downarrow\uparrow}=V_m^{\downarrow\uparrow,\uparrow\downarrow}$ and  $V_m^{\times}\equiv V_m^{\uparrow\downarrow,\uparrow\downarrow}=V_m^{\downarrow\uparrow,\downarrow\uparrow}$. In terms of these pseudopotentials, the interaction Hamiltonian reads:
\begin{align}
\hat V =& \sum_M \Bigg[ \sum_{m {\rm \ odd}} \left( V_m^{\uparrow}  \ket{mM,\uparrow\uparrow}\bra{mM,\uparrow\uparrow} +  V_m^{\downarrow} \ket{mM,\downarrow\downarrow}\bra{mM,\downarrow\downarrow} \right) + \nonumber \\
& \sum_{m} V_m^{\|} \left( \ket{mM,\uparrow\downarrow}\bra{mM,\uparrow\downarrow} + \ket{mM,\downarrow\uparrow}\bra{mM,\downarrow\uparrow} \right) + \sum_{m} V_m^{\times} \left( \ket{mM,\uparrow\downarrow}\bra{mM,\downarrow\uparrow} + \ket{mM,\downarrow\uparrow}\bra{mM,\uparrow\downarrow} \right) \Bigg].
\label{int}
\end{align}

There are two main differences to conventional bilayer (or spin) systems: First, there are two different intra-level pseudopotentials. This breaks ${\cal Z}_2$ symmetry present in systems of equivalent layers.  Second, the inter-level interactions do not only consist of density-density-interactions, $V_m^{\|}$, but also contain exchange interactions, $V_m^{\times}$, usually not present in bilayer or spin systems. Regarding the first difference we note that, as seen in Fig.~2(a) in the main text, the different intra-level pseudopotentials differ strongly only at $m=0$. Since only odd values of $m$ contribute to the fermionic system, we expect only a weak effect of this ${\cal Z}_2$ symmetry breaking. 

In order to capture the role of the exchange interactions, we introduce a spin basis in terms of singlet and triplet configurations:
\begin{align}
\ket{+} &= \frac{1}{\sqrt{2}} \left( \ket{\uparrow\downarrow} + \ket{\downarrow\uparrow} \right), \nonumber \\ 
\ket{-} &= \frac{1}{\sqrt{2}} \left( \ket{\uparrow\downarrow} - \ket{\downarrow\uparrow} \right). \nonumber
\end{align}
Re-writing Eq. (\ref{int}) in this basis, we get
\begin{align}
\hat V =& \sum_M \Bigg[ \sum_{m {\rm \ odd}} V_m^{\uparrow} \left( \ket{mM,\uparrow\uparrow}\bra{mM,\uparrow\uparrow} +  V_m^{\downarrow} \ket{mM,\downarrow\downarrow}\bra{mM,\downarrow\downarrow} \right) + \nonumber \\
&  \sum_{m {\rm \ odd}} \left[ V_m^{\|} + V_m^\times \right] \ket{mM,+}\bra{mM,+} + \sum_{m {\rm  \ even}} \left[ V_m^{\|} - V_m^\times \right] \ket{mM,-}\bra{mM,-}  \Bigg].
\end{align}
We see that symmetry demands to the wave function allow to give up the distinction between $V_m^{\|}$ and $V_m^{\times}$ if we define the inter-level interaction as
\begin{align}
V_m^{\rm inter} = \begin{cases}
	V_m^{\|} + V_m^\times &\text{if $m$ is odd,}\\
	V_m^{\|} - V_m^\times &\text{if $m$ is even.}
\end{cases}
\end{align}
This allows to directly compare the inter-level interactions in Eq. (\ref{int}) with models characterized by a single inter-layer interaction (i.e. models relevant for bilayer or spin systems). As seen in Fig.~2(b), $V_1^{\rm inter}$ rather than $V_0^{\rm inter}$ becomes the dominant contribution, when the first and the second graphene LL are coupled. As we have shown by explicit numerics in the main text, this will result in the formation of singlet ground states, or even of quantum Hall phases which are derived from a hollow-core model (i.e. $V_m^{\rm inter} \propto \delta_{m,1}$ and $V_m^{\rm intra} \propto \delta_{m,1}$), like the Haldane-Rezayi phase.

\section{Forms of the trial wave functions}
In this section we briefly review the form of the trial wave functions considered in this work for both $\nu=1/2$ and $\nu=2/3$ fillings. The simplest two component wave functions belong to Halperin (m,m,n) family \cite{Halperin83} and have Abelian excitations:
\begin{equation}
\Psi_{(m,m,n)}\left(\{z_i\}, \{w_i\}\right)=\prod_{i<j}\left(z_i-z_j\right)^m\prod_{i<j}\left(w_i-w_j\right)^m\prod_{i,j}\left(z_i-w_j\right)^n,
\end{equation}
where $z_i$ and $w_i$ are complex coordinates of the electrons for two components, $i=1\dots\frac{N}{2}$ and Gaussian factor $\exp\left[-\sum_i\left(\left|z_i\right|^2+\left|w_i\right|^2\right)/4l^2\right]$ is implicitly assumed in all formulas in this section. For filling $1/2$ the candidate Halperin state is (3,3,1), which using Cauchy determinant identity can also be written in paired form
\begin{equation}
\Psi_{(3,3,1)}\left(\{z_i\}, \{w_i\}\right)=\mathrm{Det}\left(\frac{1}{z_i-w_j}\right)\prod_{i<j}(x_i-x_j)^2,
\end{equation} 
where $x_i$ denotes particles in both components and index $i$ is running $1\dots N$. The (3,3,1) state is spin singlet and is zero energy ground state of the interaction Hamiltonian with only intralayer $V^\mathrm{intra}_{1}$ and interlayer $V^\mathrm{inter}_{0}$ pseudopotentials being non-zero. Another candidate wave function for $1/2$ filling, which is again spin singlet is Haldane-Rezayi(HR) \cite{HR} state which is obtained by multiplying (3,3,1) state with permanent $\mathrm{Per}\left(1/\left(z_i-w_j\right)\right)$. Using linear algebra identity HR state can be written in the following form
\begin{equation}
\Psi_\mathrm{HR}\left(\{z_i\}, \{w_i\}\right)=\mathrm{Det}\left(\frac{1}{\left(z_i-w_j\right)^2}\right)\prod_{i<j}(x_i-x_j)^2.
\end{equation}
HR state is characterized with non-Abelian excitations and is the exact zero energy eigenstate of hollow-core (only $V^\mathrm{intra}_{1}$ and $V^\mathrm{inter}_{1}$ being non-zero) interaction Hamiltonian. When $V_{0}$ pseudopotentials become comparable with $V_{1}$ the ground state of the system has larger overlap with gapped Jain's singlet phase compared to gapless HR state \cite{Moran12}. Jain's singlet has the following form \cite{Belkhir93}
\begin{equation}
	\Psi^{1/2}_\mathrm{JSS}\left(\{z_i\}, \{w_i\}\right)=\mathcal{P}_\mathrm{LLL}\Phi_{\nu=2}\left(\{x_i\}\right)\Phi_{(1,1,0)}\left(\{z_i\}, \{w_i\}\right)\Phi_{\nu=1}\left(\{x_i\}\right),
\end{equation}
where $\mathcal{P}_\mathrm{LLL}$ denotes the projection into lowest LL and $\Phi_{\nu=n}\left(\{x_i\}\right)$ denotes completely filled $n$ LLs. It can be represented in an alternative form, which reveals its $d$-wave pairing structure \cite{Moran12} 
\begin{equation}
	\Psi^{1/2}_\mathrm{JSS}\left(\{z_i\}, \{w_i\}\right)=\mathrm{Det}\left(\frac{\partial_{z_i}-\partial_{w_j}}{z_i-w_j}\right)\Phi^2_{\nu=1}\left(\{x_i\}\right).
\end{equation} 
For filling $\nu=2/3$ possible spin-singlet phases from Halperin (m,m,n) family include (1,1,2) and (3,3,0) states. The case when interlayer correlation is stronger than intralayer does not correspond to homogeneous state \cite{Gail08}, therefore (1,1,2) state can be excluded. The (3,3,0) state corresponds to two copies of $1/3$ Laughlin states without interlayer correlation and is the zero energy eigenstate of the interaction Hamiltonian with only $V^\mathrm{intra}_{1}$ being non-zero. Another potential candidate is the Jain's spin-singlet state with Abelian excitations \cite{Wu:1993gq,JainBook}
\begin{equation}
	\Psi^{2/3}_\mathrm{JSS}\left(\{z_i\}, \{w_i\}\right)=\mathcal{P}_\mathrm{LLL} \prod_{i<j}\left|z_i-z_j\right|^2\left|w_i-w_j\right|^2\Psi^\ast_{(1,1,2)}\left(\{z_i\}, \{w_i\}\right).
\end{equation}
This is the singlet phase observed for usual bilayer system with small interlayer tunneling and layer separation \cite{McDonald96,Geraedts15}. There are also several exotic phases proposed for $2/3$ filling which support non-Abelian excitations and are possible candidates for spin-singlet phase observed in this work. The simplest one is the interlayer Pfaffian \cite{ardonne02,maissam10}
\begin{equation}
	\Psi_\mathrm{inter}\left(\{z_i\}, \{w_i\}\right)=\mathrm{Pf}\left(\frac{1}{x_i-x_j}\right)\Psi_{(2,2,1)}\left(\{z_i\}, \{w_i\}\right),
\end{equation}
where $\mathrm{Pf}$ denotes the Pfaffian of antisymmetric matrix and corresponds to p-type pairing between all particles. On the sphere this is the zero energy eigenstate of the following interaction Hamiltonian
\begin{equation}
	H=\sum_{i<j<k}V_0P_{ijk}\left(\frac{3}{2}N_{\Phi}-3,\frac{3}{2}\right)+V_1P_{ijk}\left(\frac{3}{2}N_{\Phi}-2,\frac{1}{2}\right)+V_2P_{ijk}\left(\frac{3}{2}N_{\Phi}-1,\frac{1}{2}\right),
\end{equation}
where $P_{ijk}(L,S)$ denotes the three particle projection operator into the state with total angular momentum $L$ and total spin $S$, $N_\Phi$ is the number of flux quantum threading the sphere. For $\nu=2/3$ $N_\Phi=3N_e/2-3$, where $N_e$ is the number of electrons and $\mathcal{S}=3$ denotes the shift. The next one is the intralayer Pfaffian state \cite{maissam10}
\begin{equation}
	\Psi_\mathrm{intra}\left(\{z_i\}, \{w_i\}\right)=\mathrm{Pf}\left(\frac{1}{z_i-z_j}\right)\mathrm{Pf}\left(\frac{1}{w_i-w_j}\right)\Psi_{(2,2,1)}\left(\{z_i\}, \{w_i\}\right),
\end{equation}
and this corresponds to p-type pairing between the particles in each layer. Based on that this phase is realizable when number of particles $N_e$ is divisible by 4. The third possible candidate is the bilayer Fibonacci phase \cite{vaezi14} which is a state based on $\mathrm{SU(3)_2}$ Chern-Simons theory, some of the quasiparticles obeying non-Abelian statistics. It was identified that hollow core interaction Hamiltonian most likely supports the bilayer Fibonacci phase \cite{LiuVaezi15}. 
Finally, as a spin polarized state for large interlayer tunneling we have considered particle-hole conjugate of 1/3 Laughlin state
\begin{equation}
	\Psi_\mathrm{P-H}=\mathcal{P}_\mathrm{LLL}\prod_{i<j}\left(z_i-z_j\right)^2\Phi_{\nu=-2}\left(\{x_i\}\right),
\end{equation}
where $\Phi_{\nu=-2}\left(\{x_i\}\right)=\Phi^\ast_{\nu=2}\left(\{x_i\}\right)$ is the wave function of $\nu=-2$ integer quantum Hall state. As the Laughlin state this is the zero energy eigenstate of the interaction Hamiltonian with only $V_1$ pseudopotential being non-zero.
\putbib
\end{bibunit}
\end{document}